\documentclass[a4paper,11pt,twoside]{article}
\pdfoutput=1
\usepackage{latexsym}
\usepackage{amsmath}
\usepackage{amssymb}
\usepackage{graphicx}

\begin{document}

\title{\textbf{Investigation of the maximum amplitude increase from the Benjamin--Feir instability}}

\author{\textbf{Natanael Karjanto\thanks{E-mail : n.karjanto@math.utwente.nl},
E. van Groesen\thanks{E-mail : groesen@math.utwente.nl}, and Pearu
Peterson\thanks{On leave from Center of Nonlinear Studies,
Institute of Cybernetics at Tallinn Technical University,
Akadeemia Road 21, 12618 Tallinn, Estonia, e-mail :
pearu@cens.ioc.ee} \vspace{0.2in}} \\
Department of Applied Mathematics, University of Twente,\\
P.O. Box 217, 7500 AE Enschede, The Netherlands}
\date{}
\maketitle

\begin{abstract}
The Nonlinear Schr\"{o}dinger (NLS) equation is used to model
surface waves in wave tanks of hydrodynamic laboratories. Analysis
of the linearized NLS equation shows that its harmonic solutions
with a small amplitude modulation have a tendency to grow
exponentially due to the so-called Benjamin--Feir instability. To
investigate this growth in detail, we relate the linearized
solution of the NLS equation to a fully nonlinear, exact solution,
called soliton on finite background. As a result, we find that in
the range of instability the maximum amplitude increase is finite
and can be at most three times the initial amplitude.

\textbf{Keywords} : Nonlinear Schr\"{o}dinger equation,
Benjamin--Feir instability, soliton on finite background, maximum
amplitude increase.
\end{abstract}

\section{Introduction}

This is an initial work on 'Extreme Waves in Hydrodynamics
Laboratory'. Extreme waves here refer to very high amplitude,
steep, waves that can appear suddenly from a relatively calm sea.
Although extreme waves are very rare and unpredictable, they are
still very dangerous to ships in case they meet.

We model the problem of extreme waves using dispersive wave modes.
The specific property of dispersion is that waves with different
wave length propagate with different phase velocities. In the
following we assume that the wave field has a frequency spectrum
that is localized around one frequency. Then the envelope of the
wave field is described by the NLS equation \cite{LD94}. The NLS
equation is an amplitude equation for describing the change of
envelope of a wave group. This equation is very instrumental in
understanding various nonlinear wave phenomena : it arises in
studies of unidirectional propagation of wave packets in a energy
conserving dispersive medium at the lowest order of nonlinearity
\cite{AS99}.

In this paper we analyze the behavior of the wave group envelope
using both linear and nonlinear theories. Linear theory predicts
exponential growth of the amplitude when certain conditions are
satisfied---the Benjamin--Feir instability \cite{AS99}. However,
when the amplitude becomes large nonlinear effects must be taken
into account, that, as it turns out, will prevent further
exponential growth. The aim of this paper is to find the maximum
amplitude of waves when the amplitude growth is triggered by the
Benjamin--Feir instability. Also, we investigate how this maximum
amplitude depends on the growth rate parameter from the
Benjamin--Feir instability.

\section{Modelling of Waves Envelope}

\subsection{Linear Theory}

In linear theory of water waves, we can restrict the analysis of a
surface elevation $\eta(x,t)$ to \textit{one-mode solution} of the
form $\eta(x,t) = a\,e^{i(k x - \omega t)} + c.c.$ (complex
conjugate), where $a$ is a constant amplitude, $k$ is wavenumber,
and $\omega$ is frequency. Then a general solution is a
superposition of one-mode solutions. For surface waves on water of
constant depth $h$, the parameters $k$ and $\omega$ are related by
the following \textit{linear dispersion relation} \cite{LD94}:
\begin{equation}
\omega^{2} = g\,k\,\tanh k\,h,
\end{equation}
where $g$ is gravitational acceleration. We also write the linear
dispersion relation as $\omega \equiv \Omega(k) =
k\,\sqrt{\frac{g\,\tanh k\,h}{k}}$. This dispersion relation can
be derived from the linearized equations of the full set of
equations for water waves. The \textit{phase velocity} is defined
as $\frac{\Omega(k)}{k}$ and the \textit{group velocity} is
defined as $\frac{d\Omega}{dk} = \Omega'(k)$.

Since the linear wave system has elementary solutions of the form
$e^{\,i\,(k\,x - \Omega(k)\,t)}$, it is often convenient to write
the general solution of an initial value problem as an integral of
its Fourier components \cite{AS99} :
\begin{equation}
\eta(x,t) :=
\int^{\,\infty}_{\!\!-\infty}\!\!\alpha(k)\:e^{\,i\,(k\,x -
\Omega(k)\,t)} \,dk,      \label{Fourier}
\end{equation}
where $\alpha(k)$ is the Fourier transform of $\eta(x,0)$. Writing
dispersion relation $\Omega(k)$ as a power series (Taylor
expansion) about a fixed wavenumber $k_{0}$ and neglecting
$\cal{O}$$(\kappa^{3})$ terms, we find
\begin{equation}
\beta\,\kappa^{2} = \Omega(k_{0}+\kappa) - \Omega(k_{0}) -
\Omega'\!(k_{0})\,\kappa,
\end{equation}
where $\beta = \frac{1}{2}\,\Omega^{''}\!(k_{0})$. Let us define
$k = k_{0} + \kappa$, $\tau = t$, and $\xi = x -
\Omega'\!(k_{0})\,t$. Then equation (\ref{Fourier}) can be written
like
\begin{equation}
\eta(x,t) = e^{\,i[k_{0}\,x - \Omega(k_{0})\,t]}\!\!
\int^{\,\infty}_{\!\!-\infty}\!\!\alpha(k_{0} +
\kappa)\:e^{\,i\,\kappa\,\xi}\,e^{-i\,\beta\,\kappa^{2}\,\tau}
\,d\kappa. \label{eta}
\end{equation}
Denoting the integral in (\ref{eta}) with $\psi(\xi,\tau)$, we
find that $\psi(\xi,\tau)$ satisfies
\begin{equation}
i\,\frac{\partial \psi}{\partial \tau} + \beta\,\frac{\partial^{2}
\psi}{\partial \xi^{2}} = 0. \label{linschro}
\end{equation}
This is the \textit{linear Schr\"{o}dinger} equation for
\textit{narrow--banded} spectra. Equation (\ref{linschro}) is a
partial differential equation that describes time evolution of the
envelope of a linear wave packet \cite{AS99}. Equation
(\ref{linschro}) has a monochromatic mode solution $\psi(\xi,\tau)
= e^{i\,(\kappa\, \xi\, - \,\nu\, \tau )}$ where $\nu =
\beta\,\kappa^{2}$.

\subsection{Nonlinear Theory}

Assuming narrow-banded spectra, we consider the wave elevation in
the following form $\eta(x,t) = \psi(\xi,\tau)\, e^{i(k_{0} x -
\omega_{0} t)} + c.c.$, where $\tau = t$, $\xi = x -
\Omega'\!(k_{0})\,t$, $\psi(\xi,\tau)$ is a complex valued
function (called the \textit{complex amplitude}), $k_{0}$ and
$\omega_{0}$ are central wavenumber and frequency, respectively.
The evolution of the wave elevation is a weakly nonlinear
deformation of a nearly harmonic wave with the fixed wavenumber
$k_{0}$. If we substitute $\eta$ to the equations which describes
the physical motion of the water waves (see below), then one finds
that the complex amplitude $\psi(\xi,\tau)$ satisfies the
\textbf{\textit{nonlinear Schr\"{o}dinger (NLS) equation}}.

As example, the NLS equation can be derived from the modified KdV
equation $\eta_{t} + i\,\Omega\,(-i\partial_{x})\eta +
\frac{3}{4}\,\partial_{x}\eta^2 = 0$ \cite{EC02}. With $\tau = t$
and $\xi = x - \Omega'\!(k_{0})\,t$, the corresponding NLS
equation reads
\begin{equation}
\bigskip i\frac{\partial \psi}{\partial \tau}+\beta\,\frac{\partial
^{2}\psi}{\partial \xi^{2}%
}+\gamma\,\left| \psi\right| ^{2}\psi = 0, \quad \beta, \gamma \in
\mathbb{R}, \label{generalNLS}
\end{equation}
where $\beta$ and $\gamma$ depend only on $k_{0}$ :
\begin{eqnarray}
\beta = \frac{1}{2}\,\Omega^{''}\!\!(k_{0}),
\end{eqnarray}
\begin{equation}
\gamma = - \frac{9}{4}\,k_{0}\,\left(\frac{1}{\Omega^{'}\!(k_{0})
- \Omega^{'}\!(0)} + \frac{k_{0}}{2\,\Omega(k_{0}) -
\Omega(2k_{0})} \right),
\end{equation}
where $\omega = \Omega(k)$ is the linear dispersion relation
\cite{EG98}. The coefficients $\beta$ and $\gamma$ in this paper
have opposite signs compared to the corresponding coefficients in
\cite{EG98}. This equation arose as a model for packets of waves
on deep water.

There are two types of NLS equations :
\begin{itemize}
\item If $\beta$ and $\gamma$ have the same sign, i.e.
$\beta\,\gamma > 0$, then (\ref{generalNLS}) is called the
\textit{focusing} NLS equation (an attractive nonlinearity,
modulationally unstable [Benjamin--Feir instability]) \cite{SS99}.
\item If $\beta$ and $\gamma$ have different sign, i.e.
$\beta\,\gamma < 0$, then (\ref{generalNLS}) is called the
\textit{de-focusing} NLS equation (a repulsive nonlinearity,
stable solution) \cite{SS99}.
\end{itemize}

The wavenumber $k_{\textmd{\scriptsize{crit}}}$, for which
$\beta\,\gamma = 0$ holds, is called the \textit{critical
wavenumber} or the \textit{Davey--Stewartson value}. Using the
physical quantity $g = 9.8 $ m/s$^{2}$ and the water depth $h = 5$
m, then for wavenumbers $k_{0}
> k_{\textmd{\scriptsize{crit}}} = 0.23$, the product
$\beta\,\gamma > 0$, and the NLS equation is of focusing type. The
corresponding critical wavelength is 27.41 m.

In the following, we consider only focusing NLS equations. The NLS
equation (\ref{generalNLS}) has a \textit{plane--wave} solution
\begin{equation}
A(\tau) = r_{0}\,e^{\,i\,\gamma\,r_{0}^{2}\,\tau}.
\label{solutiondependontime}
\end{equation}
In physical variables, the surface wave elevation is given as
$\eta(x,t) = 2\,r_{0} \cos\,(k_{0}x - \omega_{0}t + \gamma r_{0}^2
t)$. Note that the corresponding phase velocity is
$\frac{\omega_{0} - \gamma\,r_{0}^2}{k_{0}}$. In the following
section, we analyze the stability of this \textit{plane--wave}
solution.

\section{Benjamin--Feir Instability} \label{BF}

To investigate the instability of the NLS plane--wave solution, we
perturb the $\xi$--independent function $A(\tau)$ with a small
perturbation of the form $\epsilon(\xi,\tau) =
A(\tau)\,B(\xi,\tau)$. We look for the cases where under a small
perturbation the amplitude of the plane wave solution grows in
time \cite{LD94}:
\begin{equation}
\psi(\xi,\tau) = A(\tau)[1 + B(\xi,\tau)]. \label{perturbation}
\end{equation}
Substituting (\ref{perturbation}) into (\ref{generalNLS}), and
ignoring nonlinear terms we obtain the \textit{linearized} NLS
equation :
\begin{equation}
i B_{\tau} + \beta\,B_{\xi\xi} + \gamma\,r_{0}^{2}(B + B^{*}) = 0.
\label{linearNLS}
\end{equation}
We seek the solutions of (\ref{linearNLS}) in the form
\begin{equation}
B(\xi,\tau) = B_{1}e^{(\sigma \tau + i \kappa \xi)} +
B_{2}e^{(\sigma^{*}\tau - i \kappa \xi)},
\label{perturbationsolution}
\end{equation}
where $B_{1}, B_{2} \in \mathbb{C}$, $k_{0} + \kappa$ is local
wavenumber, $\kappa$ is modulation wavenumber, and $\sigma \in
\mathbb{C}$ is called the \textit{growth rate}. If Re\,($\sigma) >
0$, then the perturbed solution of the NLS equation grows
exponentially. This is the criterion for the so--called
\textbf{Benjamin--Feir instability} of a one--wave mode with
modulation wavenumber $\kappa$ \cite{LD94}.

\begin{figure}[h]
\centering
\includegraphics[bb = 42 185 554 583,width=0.5\textwidth]{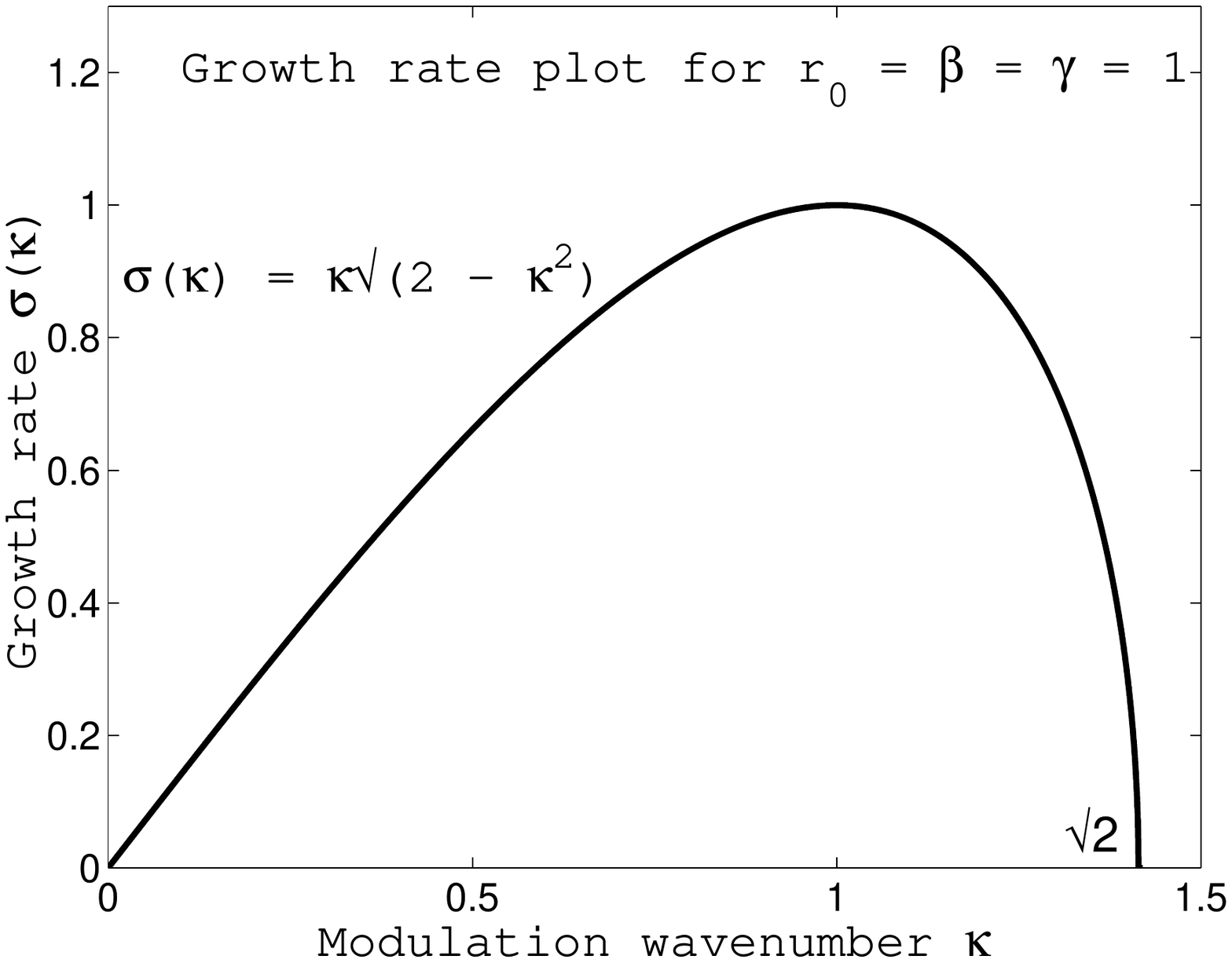}
\caption{Growth rate $\sigma$ as a function of modulation
wavenumber $\kappa$ for $r_{0} = \beta = \gamma = 1$.}
\label{plotomega}
\end{figure}

Substituting the function $B(\xi,\tau)$ in
(\ref{perturbationsolution}) into (\ref{linearNLS}) yields a pair
of coupled equations that can be written in matrix form as follows
\begin{equation}
\left (
\begin{array}{cc}
i \sigma - \beta\,\kappa^{2} + \gamma\,r_{0}^{2}  &  \gamma\,r_{0}^{2} \\
\gamma\,r_{0}^{2} & -i \sigma - \beta\,\kappa^{2} +
\gamma\,r_{0}^{2}
\end{array}
\right )
\left(
\begin{array}{c}
B_{1} \\ B_{2}^{*}
\end{array}
\right) =
\left(
\begin{array}{c}
0 \\ 0
\end{array}
\right). \label{matrix}
\end{equation}
Nontrivial solution to (\ref{matrix}) can exist only if the
determinant of the left hand side matrix is zero. This condition
reads as follows
\begin{equation}
\sigma^2 = \beta\,\kappa^{2}(2\,\gamma\,r_{0}^{2} -
\beta\,\kappa^{2}).
\end{equation}
We have the following cases :
\begin{itemize}
\item  The growth rate $\sigma$ is real and positive if
$\kappa^{2} < 2\frac{\gamma}{\beta}\,r_{0}^{2}$. This corresponds
to \textbf{Benjamin--Feir instability}. For specified values of
$\kappa$, the perturbation amplitude is exponentially amplified in
time \cite{LD94}. \item  The growth rate $\sigma$ is purely
imaginary if $\kappa^{2} > 2\frac{\gamma}{\beta}\,r_{0}^{2}$. This
corresponds to a plane--wave solution that has bounded amplitude
for all time \cite{LD94}.
\end{itemize}
Thus, the range of instability is given by
\begin{equation}
0 < |\kappa| < \left|\kappa_{\textmd{\scriptsize{crit}}}\right| =
\sqrt{\frac{2\,\gamma}{\beta}}\,r_{0}.
\end{equation}
It is easy to find that the 'strongest' instability occurs at
$\kappa_{\textmd{\scriptsize{max}}} =
\sqrt{\frac{\gamma}{\beta}}\,r_{0},$ where the maximum growth rate
is $\sigma_{\textmd{\scriptsize{max}}} = \gamma\,r_{0}^2$. Figure
\ref{plotomega} shows the plot of growth rate $\sigma$ as a
function of modulation wavenumber $\kappa$ for $r_{0} = \beta =
\gamma = 1$.

We can write the solution of the linearized NLS equation as
\begin{equation}
\psi(\xi,\tau) = r_{0}\,e^{\,i\,\gamma\,r_{0}^2\,\tau} [ 1 +
e^{\sigma(\kappa)\,\tau}(B_{1} e^{i\,\kappa\,\xi} + B_{2}
e^{-i\kappa\,\xi})],
\end{equation}
with $\sigma(\kappa) = \kappa\,\sqrt{2\,\beta\,\gamma\,r_{0}^{2} -
\beta^{2}\,\kappa^{2}}$. Since this solution is obtained from the
linearized NLS equation, it is only valid if amplitudes are small.
When the time is increased, the amplitude increases exponentially
and the linearized theory becomes invalid. Therefore, we cannot
use the solution from the linearized equation to investigate the
behavior of the maximum amplitude increase due to the
Benjamin--Feir instability. Fortunately, there exists an exact
solution to the NLS equation that describes the exact behavior of
the wave profile and that corresponds to the Benjamin--Feir
instability.

\section{Modulational Instability}

\begin{figure}[h]
\centering 
\includegraphics[bb = 43 189 570 615,width=0.7\textwidth]{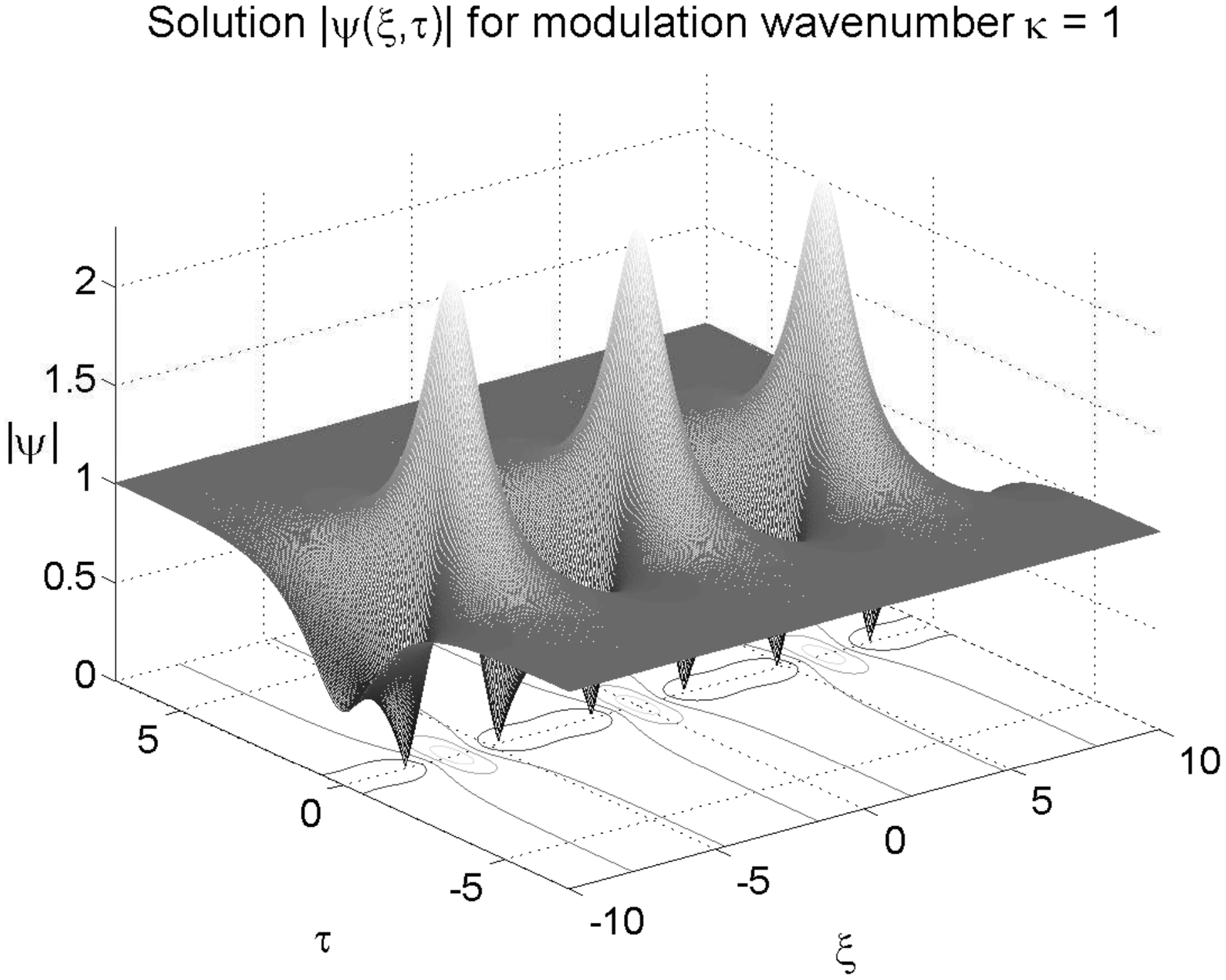}
\caption{Plot of $\left|\psi \right|$ as a function of $\xi$ and
$\tau$ for modulation wavenumber $\kappa = 1.$} \label{plot3d}
\end{figure}

In this section we investigate the relation between the maximum
amplitude of a certain solution of the NLS equation and the
modulation wavenumber $\kappa$ in the instability interval. For
simplicity, we choose the amplitude $r_{0} = 1$, and the
coefficients $\beta = \gamma = 1$. An exact solution, the
so--called \textit{soliton on finite background}, in short SFB,
(sometimes also called \textit{the second most important
solution}), of the NLS equation is given by \cite{AA97} :
\begin{equation}
\psi(\xi,\tau) := \frac{(\kappa^{2} - 1) \cosh
(\sigma(\kappa)\,\tau) + \sqrt{\frac{2-\kappa^{2}}{2}}\cos (\kappa
\xi) + i \sigma(\kappa) \sinh (\sigma(\kappa)\,\tau)} {\cosh
(\sigma(\kappa)\,\tau) - \sqrt{\frac{2-\kappa^{2}}{2}} \cos
(\kappa \xi)}\; e^{\,i\,\tau}, \label{exact2}
\end{equation}
where $0 < \kappa < \sqrt{2}$ and $\sigma(\kappa) = \kappa \sqrt{2-\kappa^{2}}$. \\
For $\kappa = 1,$ we have
\begin{equation}
\psi(\xi,\tau) := \frac{\cos \xi + i \sqrt{2} \sinh \tau
}{\sqrt{2} \cosh \tau - \cos \xi}\;e^{\,i\,\tau}. \label{exact1}
\end{equation}
Figure \ref{plot3d} shows the plot of $\left|\psi \right|$ from
(\ref{exact1}) as a function of $\xi$ and $\tau$. Note that
$\psi(\xi,\tau)$ is a $2\pi$--periodic function with respect to
$\xi$ variable.

In the following, we analyze the solution (\ref{exact2}) in
detail.
The behavior for this SFB as $\tau \rightarrow \pm \infty$ is
given by
\begin{equation}
\lim_{\tau \rightarrow \infty} \; \left|\psi(\xi,\tau)\right| =
\sqrt{(\kappa^2 -1)^2 + \kappa^2\,(2 - \kappa^2)} = 1 .
\end{equation}
Because of this property the solution (\ref{exact2}) is called as
SFB. For a 'normal' soliton, the elevation vanishes at infinity :
the 'normal' soliton is exponentially confined. For SFB, the
solution is a similar elevation on top of the finite (nonzero)
background level, here the normalized value 1.

Write the solution in the form $\psi(\xi,\tau) =
u(\xi,\tau)\,e^{\,i\,\tau}$, where $u(\xi,\tau)$ where $u$
describes the amplitude and the exponential part expresses
oscillations in time. Let us investigate the behavior of
$u(\xi,\tau)$ in time. For that consider the limiting behavior of
$\partial_{\tau}u$ as $\tau \rightarrow \pm \infty$. We have
\begin{equation}
\frac{\partial u}{\partial \tau}(\xi,\tau) =
\frac{i\sigma^{2}(\kappa) -
\sigma(\kappa)\,\sqrt{\frac{2-\kappa^2}{2}} \cos (\kappa
\,\xi)\left[\kappa^2 \,\sinh \left(\sigma(\kappa)\,\tau \right) +
i\,\sigma(\kappa)\,\cosh \left(\sigma(\kappa)\,\tau\right)
\right]}{\left[\cosh \left(\sigma(\kappa)\,\tau\right) -
\sqrt{\frac{2-\kappa^2}{2}} \cos
\left(\kappa\,\xi\right)\right]^{2}}.
\end{equation}
If $\xi \neq \frac{\pi}{2\,\kappa}$, then
\begin{equation}
\frac{\partial u}{\partial \tau} \approx - 2\,(\kappa^2 +
i\,\sigma(\kappa)) e^{-\sigma(\kappa)\,\tau} \quad \textmd{if} \;
\tau \gg 0,
\end{equation}
and
\begin{equation}
\frac{\partial u}{\partial \tau} \approx 2\,(\kappa^2 -
i\,\sigma(\kappa)) e^{\,\sigma(\kappa)\,\tau} \quad \textmd{if} \;
\tau \ll 0.
\end{equation}
If $\xi = \frac{\pi}{2\,\kappa}$, then
\begin{equation}
\frac{\partial u}{\partial \tau} \approx 4\,i\,\sigma^2(\kappa)
\,e^{-2\,\sigma(\kappa)\,\tau} \quad \textmd{for} \; \tau \gg 0,
\end{equation}
and
\begin{equation}
\frac{\partial u}{\partial \tau} \approx 4\,i\,\sigma^2(\kappa)\,
e^{\,2\,\sigma(\kappa)\,\tau} \quad \textmd{for} \; \tau \ll 0.
\end{equation}

\begin{figure}[h]
\centering 
\includegraphics[bb = 45 187 553 591,width=0.7\textwidth]{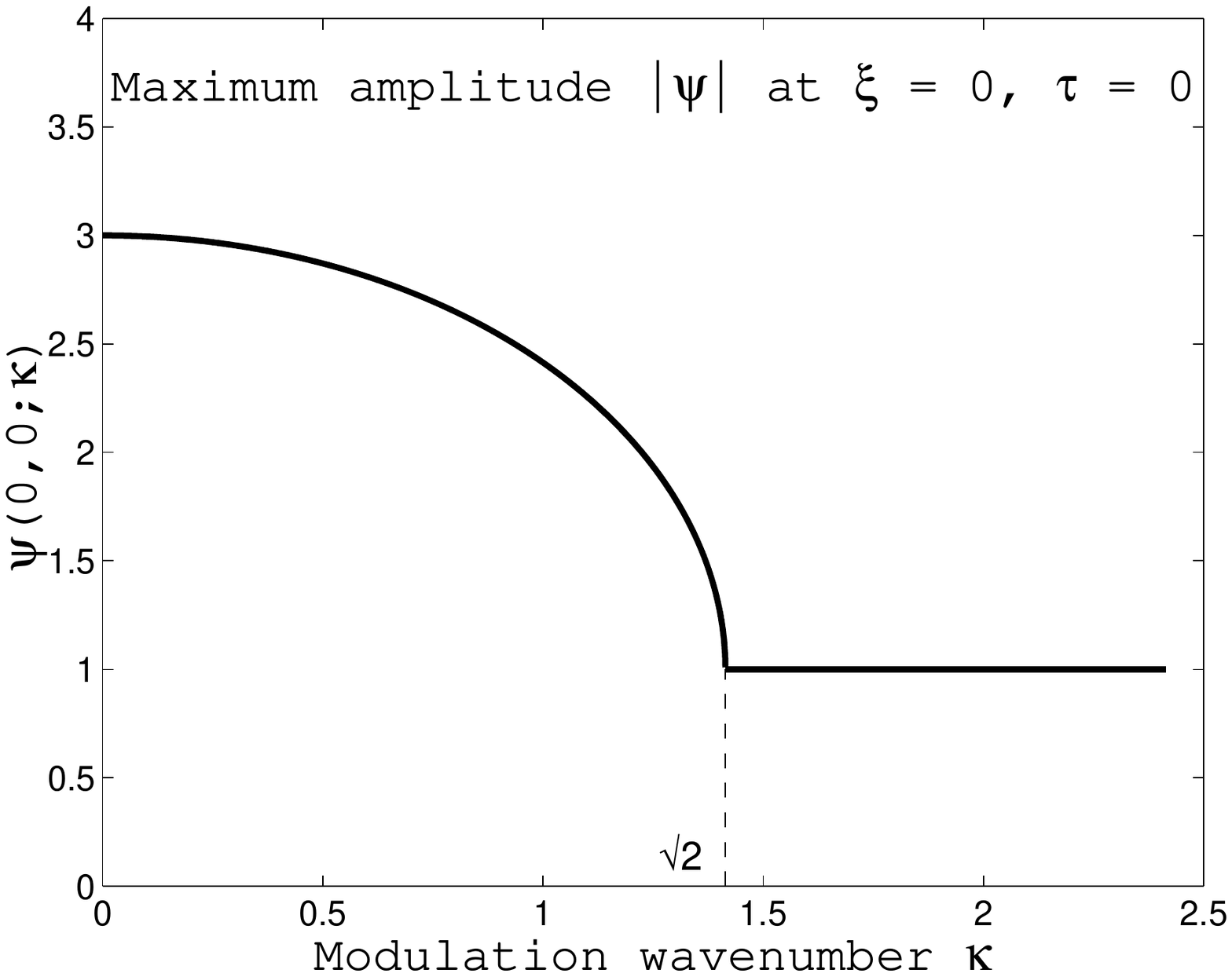}
\caption{Plot amplitude maximum $\psi(\xi=0,\,\tau=0;\kappa)$ as a
function of the modulation wavenumber $\kappa$.} \label{plotpsi}
\end{figure}
Next, let us find the relation between the maximum amplitude of
the exact solution (\ref{exact2}) with modulation wavenumber
$\kappa$, where $0 < \kappa < \sqrt {2}$. The maximum value of the
complex amplitude is at $\xi \equiv 0\;(\textmd{mod}\;2\pi)$ and
when $\tau = 0$. So, we have
\begin{equation}
|\psi|_{\textmd{\scriptsize{max}}} = |\psi(0,0\,;\,\kappa)| =
\frac{\kappa^{2} - 1 + \sqrt{1 - \frac{1}{2} \kappa^{2}}}{1 -
\sqrt{1 - \frac{1}{2} \kappa^{2}}}.
\end{equation}
Using the approximation $\sqrt{1+a} \approx 1 + \frac{1}{2} a$ for
small $a$, and apply it to our formula ($a = - \frac{1}{2}
\kappa^{2}$), we obtain
\begin{eqnarray}
\lim_{\kappa \rightarrow 0}\;|\psi(0,0;\kappa)| = 3.
\end{eqnarray}
As a result, the maximum factor of the amplitude amplification is
\begin{equation}
\lim_{\kappa \rightarrow 0}\;\frac{|\psi(0,0;\kappa)|}{\lim_{\tau
\rightarrow \pm \infty}|\psi(\xi,\tau;\kappa)|} = \frac{3}{1} = 3.
\end{equation}
Figure \ref{plotpsi} shows the plot of the maximum amplitude
$|\psi|_{\textmd{\scriptsize{max}}}$ as a function of modulation
wavenumber $\kappa$.

\begin{figure}[h]
\centering
\includegraphics[bb = 18 187 552 583,width=0.7\textwidth]{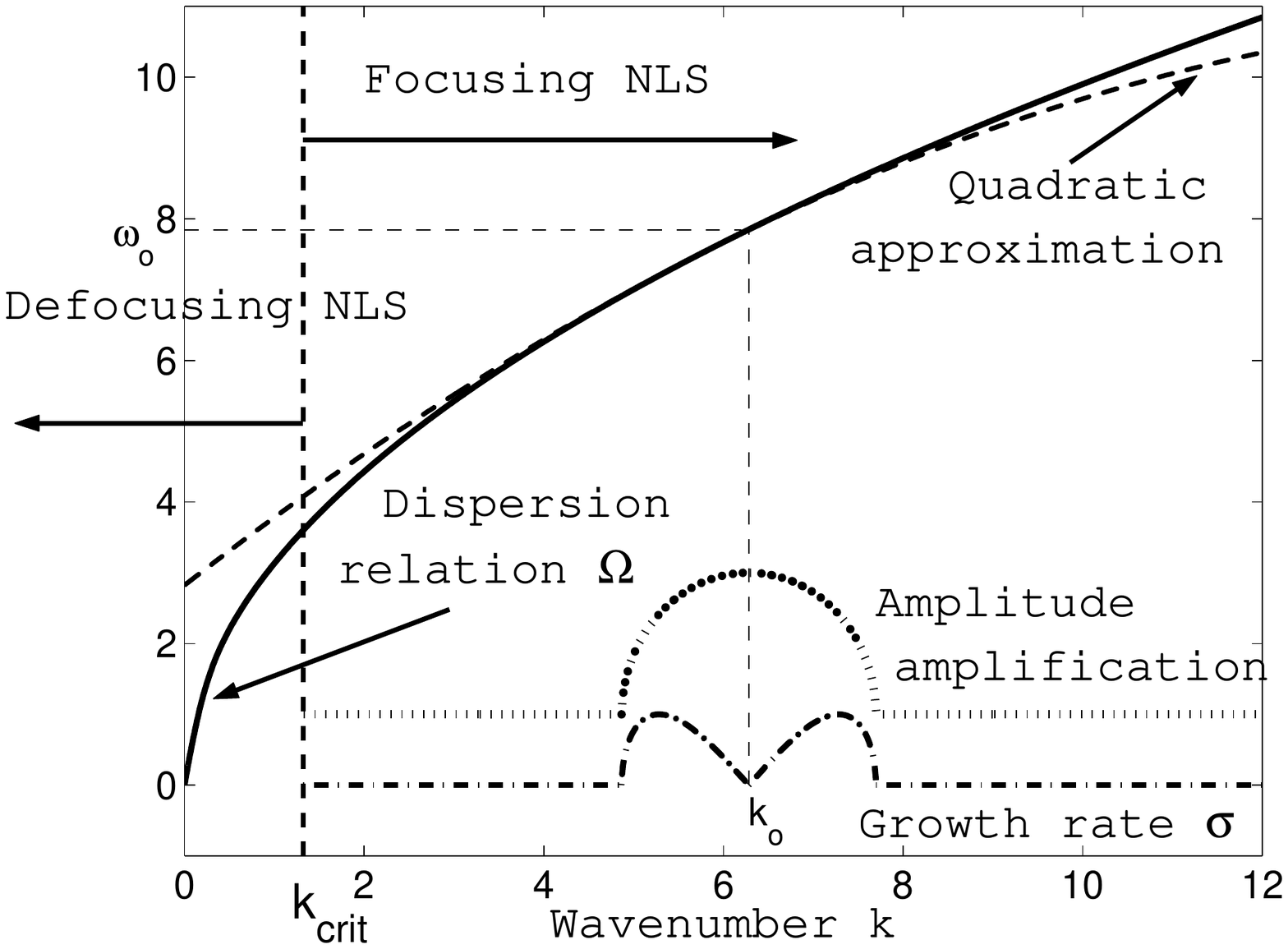}
\caption{Dispersion relation $\Omega$ and its quadratic
approximation, growth rate $\sigma$ (Benjamin--Feir instability),
and maximum amplitude of $\psi$ as functions of wavenumber $k$.
The case corresponds to $k_{0} = 2\,\pi$.} \label{plotgabung}
\end{figure}

To summarize, Figure \ref{plotgabung} compares the plot of the
dispersion relation $\Omega$ and its quadratic approximation
(leads to the NLS equation), the growth rate $\sigma$ (which is
related to Benjamin--Feir instability), and the maximum amplitude
of $\psi$ as functions of wavenumber $k$.

\section{Conclusion}

In this paper we modelled surface waves in wave tanks of
hydrodynamics laboratories using the NLS equation. We analyzed the
linearized NLS equation and obtained that its solutions have
tendency to grow exponentially. We considered also the exact
solution known as SFB of the NLS equation, that is the
continuation of this linear instability. Using this, we found the
maximum amplitude in space and time when the modulation wavenumber
is in the interval of the Benjamin--Feir instability. The main
result of this paper is that the maximum factor of the amplitude
amplification due to the Benjamin--Feir instability is three. As
we can see from Figure \ref{plotgabung}, the growth rate from the
Benjamin--Feir instability does not determine the maximum
amplitude amplification of the SFB. The results of this paper will
be used in the further study of relating the Benjamin--Feir
instability and the Phase--Amplitude equations.
\vspace{0.7cm} \newline \Large \textbf{Acknowledgement}
\vspace{0.4cm} \newline \normalsize This work is executed at
University of Twente, The Netherlands as part of the project
'Extreme Waves' TWI.5374 of the Netherlands Organization of
Scientific Research NWO, subdivision Applied Sciences STW.

\bibliographystyle{plain}
\bibliography{Karyanto}
%
%
%
%
%
%
\end{document}